\documentclass[aps,pre,preprint,showpacs,superscriptaddress]{revtex4-1}  
\usepackage{graphicx}  
\usepackage{dcolumn}   
\usepackage{bm}        
\usepackage{amssymb}   
\usepackage{times}
\usepackage{amsmath}
\usepackage{amsfonts}
\usepackage{amsthm}
\usepackage[normalem]{ulem}

\usepackage{tabularx}
\usepackage{rotating}
\usepackage{multirow}
\usepackage{array}

\usepackage{fancyhdr}
\usepackage{graphicx}
\usepackage{epsfig}
\usepackage[dvipsnames]{xcolor}
\usepackage{tikz}
\usetikzlibrary{arrows,positioning,calc}
\usetikzlibrary{matrix}
\usetikzlibrary{shapes}
\usetikzlibrary{chains}

\usepackage{lipsum}
\usepackage[english]{babel}
\usepackage[utf8]{inputenc}
\usepackage[T1]{fontenc}
\usepackage{lmodern}
\usepackage{subfigure}
\usepackage{titlesec}
\usepackage{mathtools} 
\usepackage{tcolorbox} 
\usepackage{url}
\usepackage{epstopdf}
\usepackage{bigints}
\usepackage{xpatch}
\xpatchcmd\bibsection{19}{19}{}{} 
\xpatchcmd\bibsection{\begingroup}{\vskip19pt\begingroup}{}{} 

\newcommand{\til}[1]{\widetilde{#1}}
\newcommand\ddfrac[2]{\frac{\displaystyle #1}{\displaystyle #2}}
\DeclareMathOperator\arctanh{Arctanh}
\begin{document}

\widetext

\title{Storage of energy in constrained non-equilibrium systems}

\author{Y.Zhang}\email{yzhang@ichf.edu.pl}
\affiliation{Institute of Physical Chemistry, Polish Academy of Sciences, Kasprzaka 44/52, PL-01-224 Warsaw, Poland}
\author{K.Gi\.zy\'nski}\email{kgizynski@ichf.edu.pl}
\affiliation{Institute of Physical Chemistry, Polish Academy of Sciences, Kasprzaka 44/52, PL-01-224 Warsaw, Poland}
\author{A.Macio\l ek}\email{amaciolek@ichf.edu.pl}
\affiliation{Institute of Physical Chemistry, Polish Academy of Sciences, Kasprzaka 44/52, PL-01-224 Warsaw, Poland}
\affiliation{Max-Planck-Institut f{\"u}r Intelligente Systeme, Heisenbergstr.~3, D-70569 Stuttgart, Germany}
\author{R.Ho\l yst}\email{rholyst@ichf.edu.pl}
\affiliation{Institute of Physical Chemistry, Polish Academy of Sciences, Kasprzaka 44/52, PL-01-224 Warsaw, Poland}

\date{\today}

\begin{abstract}
We study a quantity $\mathcal{T}$ defined as the energy U, stored in non-equilibrium steady states (NESS) over its value in equilibrium $U_0$, $\Delta U=U-U_0$ divided by the heat flow $J_{U}$ going out of the system. A recent study suggests that $\mathcal{T}$ is minimized in steady 
states (Phys.Rev.E.\textbf{99}, 042118 (2019)). We evaluate this hypothesis using an ideal gas system with three methods of energy delivery: from a uniformly distributed 
energy source, from an external heat flow through the surface, and from an external matter flow. 
By introducing internal constraints into the system, we determine $\mathcal{T}$ with and without constraints and find that $\mathcal{T}$ is 
the smallest for unconstrained NESS. We find that the form of the internal energy in the studied NESS follows $U=U_0*f(J_U)$. In this context, we discuss natural variables for NESS,   define the embedded energy (an analog of Helmholtz free energy for NESS), and provide its interpretation.
\end{abstract}

\maketitle

\section{Introduction}
The basis of equilibrium thermodynamics relies on the existence of the equilibrium state. The equilibrium state can be characterized by a set of appropriate parameters and some kind of energy-based function of these parameters and internal constraints. The constraints allow comparing this function in the state of equilibrium with states of constrained equilibrium \cite{holyst2012thermodynamics}. For a monoatomic system, the internal energy $U(S,V,N)$ is a function of three parameters of state, namely entropy $S,$ volume $V,$ and the number of particles $N,$ which fully characterize all thermodynamics changes that can occur in the system. For an unconstrained isolated system, $S(U,V,N)$ is maximized at constant $U, V, N$ with respect to all states obtained by internal constraints. 

A prerequisite for any system to become non-equilibrium is a continuous energy flow. This macroscopic flow of energy leads to an increase of the system energy up to the point when the energy flow into the system matches exactly the flow out of the system. At this point, the non-equilibrium steady state is reached. Two parameters characterize the NESS: the flow $J_{U}$  and the internal energy $U$. We   show that $U=U_{0}*f(J_{U})$, where $U_{0}$ is the energy at equilibrium. We   make three case studies: (i) a system internally heated between two parallel walls of the same temperature; (ii) a heat flow between two parallel plates of different temperature; and (iii) a Poisseulle flow between two parallel plates. 

Non-equilibrium states are ubiquitous in nature and truly equilibrium states are exceptions. However, despite many decades of study, we have not reached the same status of understanding of non-equilibrium  states as we have for equilibrium ones. 
There is no systematic approach for dealing with NESS. 
Attempts to create such approaches include: minimum/maximum entropy production principle \cite{martyushev2013entropy}, steady state thermodynamics \cite{oono1998steady}, and driven lattice gas systems \cite{dickman2018driven}. 
The heat flowing into the system is recognized as a source of entropy increase in \cite{bartlett2016maximum, morriss2012deterministic, holubec2017thermal}. In information theoretical techniques and extended thermodynamics, the heat flow appears as a natural thermodynamic variable in non-equilibrium steady states. The entropy of some ideal systems such as ideal gases, photons, phonons, and  ideal harmonic chains, among others, in the presence of a heat flow  is studied in \cite{jou2010extended, lebon2008understanding, luzzi2002predictive, muller2013rational, eu2013nonequilibrium}. However, the energy that has to be stored in NESS has not been recognized  as  potentially a function of state, from which in principle we could derive all properties of NESS \cite{Robert}.

In this paper, we attempt to address the latter issues. In a recent paper, a quantity
\begin{equation}
\label{eq:1}
\mathcal{T}=\frac{U-U_0}{J_U} 
\end{equation}
  is shown to be minimized in steady states for three different systems \cite{Robert}. 
This quantity has the dimension of time. In \cite{Robert}, $\mathcal{T}$ is shown to coincide with the characteristic time scale of the system energy dissipation immediately after the shutdown of external energy flow.
The minimization is demonstrated through introducing a constraint into the system and showing that $\mathcal{T}$ for the unconstrained system is always less than in the constrained system. 
In this paper, we analyze energy storage and $\mathcal{T}$ in Systems   (i)--(iii) defined above (these systems are different from the ones in \cite{Robert}) and we arrive at the same conclusions as in  \cite{Robert}. Moreover, we introduce the embedded energy, which is an analog of the Helmholtz free energy for NESS, and provide its interpretation. 

We point out that, in this paper, we extensively use the temperature profile to obtain the stored energy $U - U_{0}$. The local temperature is defined from the ideal gas law. It would be interesting, however, to consider using effective temperature in non-equilibrium systems and to study their role in energy storage \cite{casas2003temperature, puglisi2017temperature}. 
\section{Models and Results}
We consider an ideal gas driven out-of-equilibrium by three different ways of energy delivery that are common in physical realizations.
In   Case (i)  the energy is delivered through a homogeneous energy source, in   Case (ii) by an external heat flow, and in   Case (iii) by an external matter flow.

In steady states, the local energy does not change in time. Therefore, from local energy conservation, we have
\begin{equation}
\nabla \cdot \vec{J} = - k \nabla^{2}T(\vec{r}) = \sigma_{E}(\vec{r}),
\label{T_r}
\end{equation}
where $\sigma_{E}(\vec{r})$ is the local energy source at the position $\vec{r}$. Here, we   assume  the Fourier's law for the local heat flux, 
\begin{equation}
\vec{J}(\vec{r}) = - k \nabla T(\vec{r}), 
\label{J_r}
\end{equation}
where $k$ is the heat conductivity and $T(\vec{r})$ is the local temperature.
We further assume that in the NESS the ideal gas law is fulfilled locally and that the pressure (and hence the energy density) is constant. 
From these assumptions, we obtain the following relation between the energy density $\epsilon$ and the temperature profile, 
\begin{equation}
\epsilon = \dfrac{\epsilon_{0}}{T_{0}}\ddfrac{V}{\int_{V}\dfrac{d^3r}{T(\vec{r})}}, 
\label{e_intro}
\end{equation}
where $V$ is the volume of the system and  $\epsilon_{0}$ and $T_{0}$ are the energy density and temperature at equilibrium, respectively. In this paper, we denote the corresponding equilibrium value of a variable with a subscript $0$. We show the derivation of Equation (\ref{e_intro}) in Appendix A. 
From $\epsilon$, we define the stored energy as
\begin{equation}
\Delta U = U - U_{0} = (\epsilon - \epsilon_{0})V. 
\label{eq:stored_en}
\end{equation}

Without performing work (as is the case for our systems), all out-going energy flow $\dot{E}_{out}$ is in the form of heat, which we denote as $\Phi_{out}$, 
\begin{equation}
\Phi_{out} = \int\int_{S} \vec{J} \cdot \hat{n} dS, 
\end{equation}
where $S$ is the area through which the heat flows out and $\hat{n}$ is the unit normal vector.
In the steady state, the total energy flow into the system equals the total energy flow out of the system
$\dot{E}_{in}^{ss} = \dot{E}_{out}^{ss} = \Phi_{out}^{ss}$. 
In the following, we denote the out-going energy flow in the steady state by $J_U \equiv \Phi_{out}^{ss}$.

By introducing geometrical constraints, the system is partitioned into two subsystems. These constraints do  not change the local expressions for $\vec{J}(\vec{r})$ and $T(\vec{r})$. In addition, for each subsystem, definitions of the stored energy $U_{i}, i=1,2$ and the out-going heat flow $J_{U_i}, i=1,2$ remain the same. On the other hand, the subsystem energy density depends on the constrain in general. However, in all three cases, the number of particles in each subsystem is kept proportional to the volume of the system,   {i.e.}, $N_{i}/V_{i} = N/V = n_{0}$. As a result, the expression of $\epsilon_{i}$ has the same form as Equation (\ref{e_intro}) (see Appendix \ref{appendix:A}). 
For the constrained system, we define the stored energy as
\begin{equation}
\label{eq:en_st_con}
\Delta U_{tot} = \Delta(U_1+U_2) = \sum_{i} \epsilon_{i}V_{i} - \epsilon_{0}V, 
\end{equation}
and the total out-going heat flow as $J_{tot}\equiv J_{U_1}+J_{U_2}$. For every case studied in this paper, we   compare the ratio $\mathcal T_{1|2}$ for the constrained system, 
\begin{equation}
\label{eq:2}
\mathcal T_{1|2}=\frac{\Delta U_{tot}}{J_{tot}}, 
\end{equation}
with the ratio $\mathcal T$ (see Equation (\ref{eq:1})) for the unconstrained system. 

\subsection{Energy Source}

\begin{figure}[htb!]
\centering
\includegraphics[scale=0.5]{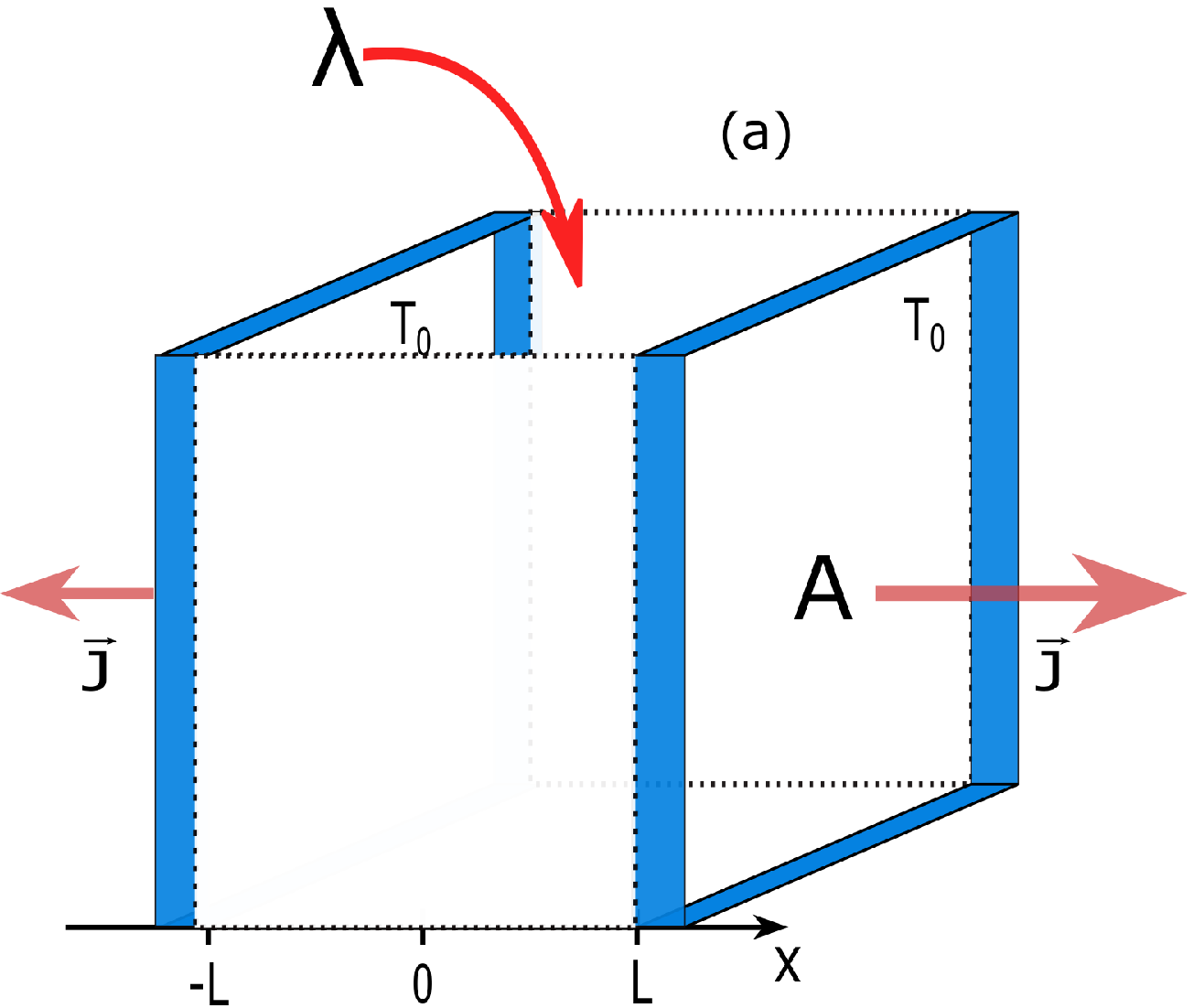}
\includegraphics[scale=0.5]{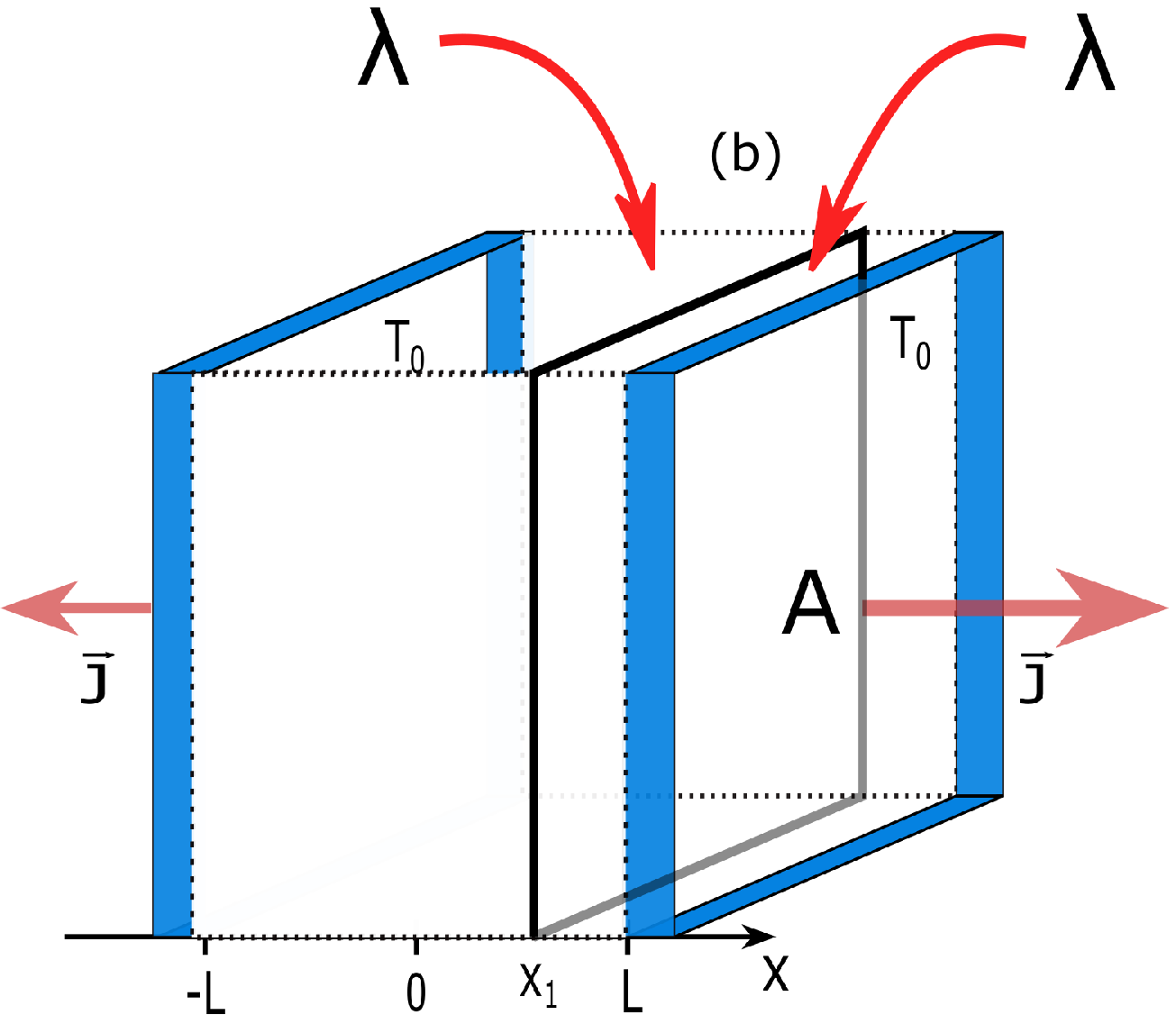}
\caption{Schemes of (\textbf{a}) unconstrained and (\textbf{b}) constrained ideal gas model under an external energy supply. The two diathermal walls of area $A$ and temperature $T_{0}$ are positioned at $x=\pm L$. An external energy is supplied homogeneously to the bulk with a density $\lambda$. The heat flux $2\vec J$ leaves the system through boundaries. In (\textbf{b}), the vertical plane at $x=x_{1}$ represents the internal constraint, which is a diathermal wall.}
\label{scheme-sigma}
\end{figure}

In   Case (i), we consider a three-dimensional ideal gas placed between two diathermal walls of area $A$ ($A \to \infty$). The walls are kept at temperature $T_{0}$ and are fixed at $x = \pm L$. The energy source is distributed homogeneously over the system with $\sigma_{E}(\vec{r}) = \lambda$. As internal constraints, we choose a diathermal wall and fix it at $x_{1} \in (-L,L)$. This wall separates the system into two subsystems $1$ and $2$ with volumes $V_{1} = A(L+x_{1})$ and $V_{2} = A(L-x_{1})$, respectively. A scheme of the system is shown in Figure \ref{scheme-sigma}. 

Consider first the unconstrained system.~As the coordinates $y$ and $z$ do not influence the temperature profile, it is sufficient to consider $x-$dependence.~The temperature profile $T(x)$ is obtained by solving Equation (\ref{T_r}), which now has the form $- k \partial_{x}^{2}T = \lambda$. 
Using dimensionless variables $\til{\lambda} = \lambda L^{2}/kT_{0}$, $\til{T}(x) = T(x)/T_{0}$ and normalizing  $x$ to $\til{x}=x/L$, we obtain 
\vspace{12pt}
\begin{equation}
\til{T}(\til{x}) = - \dfrac{\til{\lambda}}{2}\til{x}^{2} + 1 + \dfrac{\til{\lambda}}{2}. 
\end{equation}
Using Equation (\ref{e_intro}), we find the energy density to be
\begin{equation}
\epsilon = \ddfrac{\epsilon_{0}\sqrt{\til{\lambda}(\til{\lambda}+2)}}{2\arctanh(\sqrt{\til{\lambda}/(\til{\lambda}+2)})}. 
\label{dia_epsilon}
\end{equation}
As stated  above, the out-going heat flow equals   the in-coming energy flow, $\dot{E}_{in} = 2L A\lambda = J_U$. Combining with Equation (\ref{dia_epsilon}), we find 
\begin{equation}
\dfrac{\Delta U}{J_{U}} = \dfrac{\epsilon - \epsilon_{0}}{\lambda}. 
\label{dia_F}
\end{equation}

In the presence of the diathermal wall, the boundary conditions at the constraint are $T_{1}(x_{1}) = T_{2}(x_{1})$ and $d T_{1}(x)/dx|_{x_1} = d T_{2}(x)/dx|_{x_1}$.~Solving for the subsystem temperature profile with corresponding boundary conditions, we find that $T_{i}(x)$ is not changed by the constraint,   {i.e.,} $T_{1}(x) = T_{2}(x) =T(x)$, in their respective domains. 
Therefore, we obtain the energy densities as 
\begin{align}
\epsilon_{1} &= \ddfrac{\epsilon_{0}(1+\til{x_{1}})\sqrt{\til{\lambda}(\til{\lambda}+2)}}
{2\arctanh \big( \sqrt{\til{\lambda}/(\til{\lambda}+2)} \big) + 
2\arctanh \big( \til{x_{1}}\sqrt{\til{\lambda}/(\til{\lambda}+2)} \big)}, \label{eq:e1}\\
\epsilon_{2} &= \ddfrac{\epsilon_{0}(1-\til{x_{1}})\sqrt{\til{\lambda}(\til{\lambda}+2)}}
{2\arctanh \big( \sqrt{\til{\lambda}/(\til{\lambda}+2)} \big) - 
2\arctanh \big( \til{x_{1}}\sqrt{\til{\lambda}/(\til{\lambda}+2)} \big)} \label{eq:e2}. 
\end{align}
As the total energy source does not change, the out-going heat flow is not changed either, $J_{U_1}+J_{U_2} = J_U = 2LA\lambda$. 
Together with Equation (\ref{eq:e1}) and (\ref{eq:e2}), we have
\begin{equation}
\mathcal{T}_{1|2} \equiv \dfrac{\Delta U_{tot}}{J_{tot}} = \dfrac{\Delta U_{tot}}{J_{U}} = 
\dfrac{\epsilon_{1}(1+\til{x_{1}}) + \epsilon_{2}(1-\til{x_{1}}) - 2\epsilon_{0}}{2\lambda}. 
\label{dia_F12}
\end{equation}

Now, we compare Equations (\ref{dia_F}) and   (\ref{dia_F12}). The relation reduces to 
\begin{equation}
\label{comp}
\epsilon \sim \dfrac{\epsilon_{1}(1+\til{x_{1}})}{2} + \dfrac{\epsilon_{2}(1-\til{x_{1}})}{2}. 
\end{equation}
Dividing both sides of Equation (\ref{comp}) by $\epsilon$ gives 
\begin{equation}
\label{comp1}
1 \sim \dfrac{(1+\til{x_{1}})^{2}}{2+2 a} + \dfrac{(1-\til{x_{1}})^{2}}{2-2 a}, 
\end{equation}
where $a = \arctanh (\til{x_{1}}\sqrt{\til{\lambda}/(\til{\lambda}+2)}) / \arctanh(\sqrt{\til{\lambda}/(\til{\lambda}+2)})$. 
For $\til{x_{1}} \in (-1,1)$, $a \in (-1,1)$ and $(2+2a)(2-2a) \geq 0$. We multiply this by both sides of   Equation (\ref{comp1}) and rearrange the terms to obtain
\begin{equation}
0 \sim (a-\til{x_{1}})^{2}. 
\end{equation}
Since $(a-\til{x_{1}})^{2} \geq 0$, we have verified for this model
\begin{equation}
\mathcal{T} \leq \mathcal{T}_{1|2}. 
\end{equation}

\subsection{Heat Flow} 

\begin{figure}[htb!] 
\centering
\includegraphics[scale=0.6]{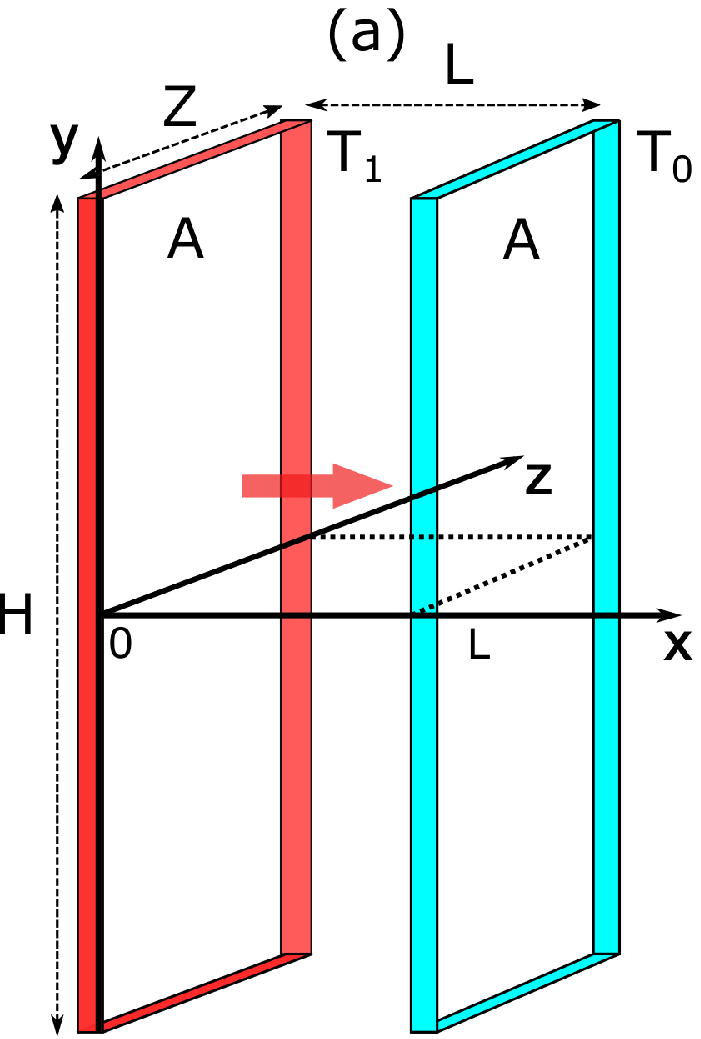}
\includegraphics[scale=0.6]{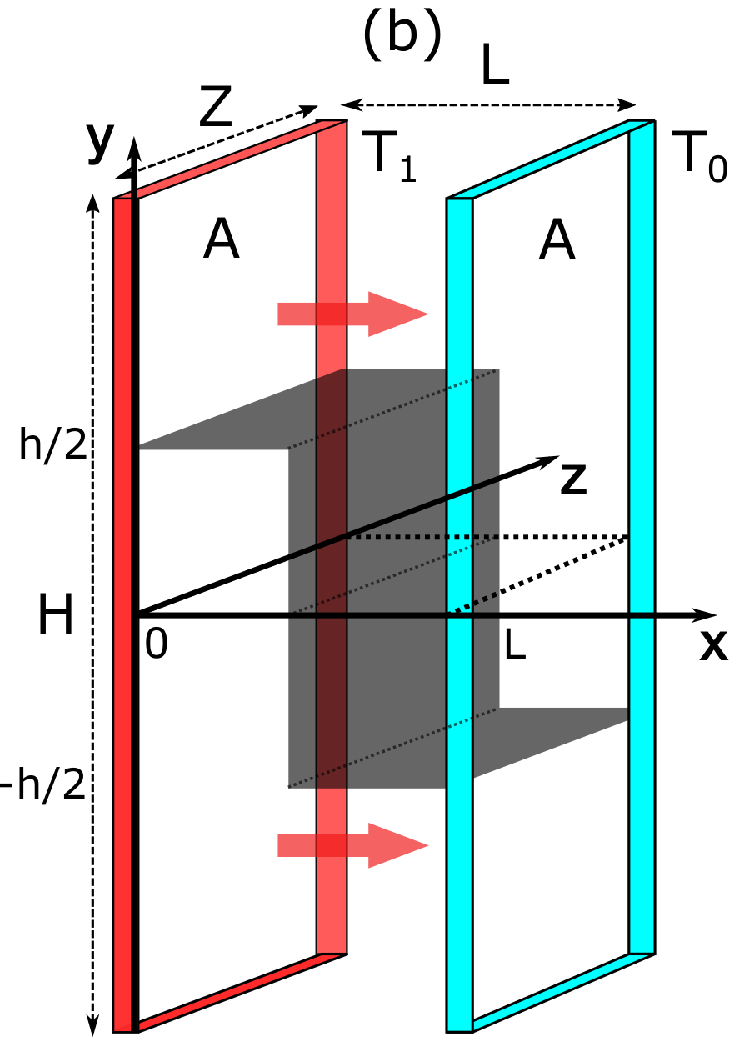}
\includegraphics[scale=0.6]{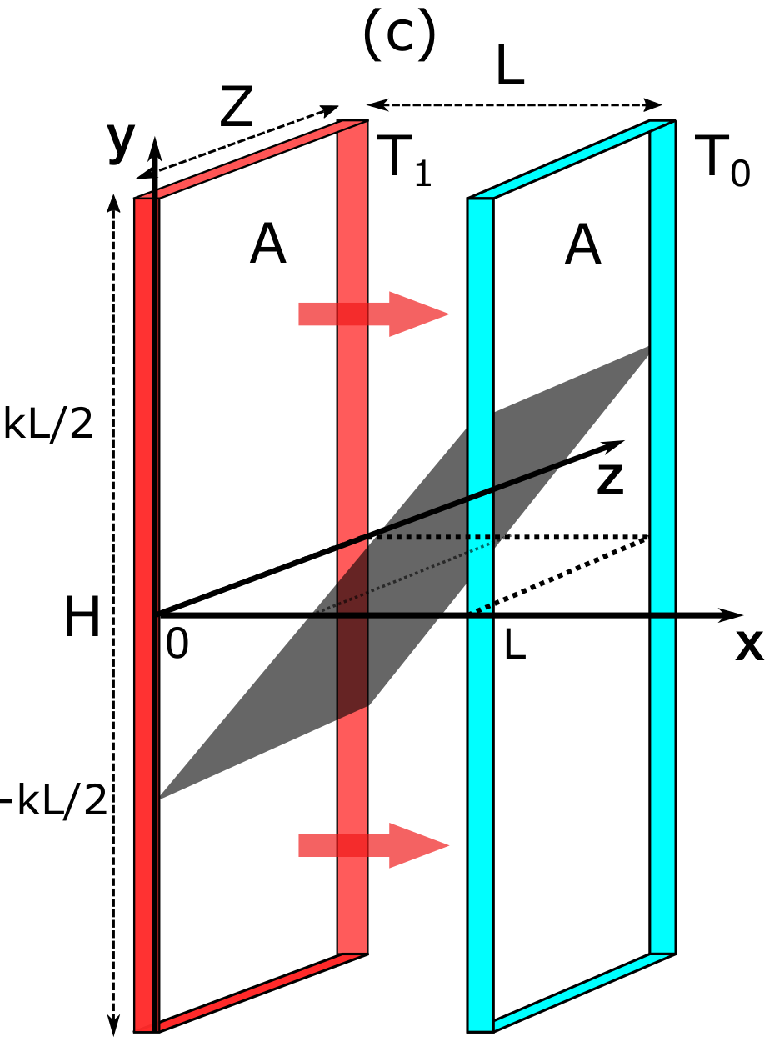}
\caption{Schemes of (\textbf{a}) unconstrained, (\textbf{b}) and (\textbf{c}) constrained ideal gas systems with an external heat flow. Two diathermal walls at temperatures $T_{1}$ and $T_{0}$ are placed at $x=0$ and $L$, respectively. 
In (\textbf{b},\textbf{c}), the black surface inside the system represents the constraint, which is an adiabatic wall. In (\textbf{b}), the constraint has a height $h$ and extends from $(0,h/2)$ to $(L/2, h/2)$ to $(L/2, -h/2)$ to $(L,-h/2)$. In (\textbf{c}), the constraint has a slope $k$ and it stretches from $(0, -k L/2)$ to $(L, k L/2)$. The red arrows denote the heat flux.}
\label{fig:scheme-heat-flux}
\end{figure}

In   Case (ii), an ideal gas is in contact with two walls at different temperatures $T_{1} \geq T_{0}$.~The~walls are of a large area $A=H \times Z$ (with height $H$ and width $Z$) and are placed at $x = 0$ and $x=L$ (see Figure \ref{fig:scheme-heat-flux}). The steady state is driven by a constant heat flow through the system with no bulk energy supply,   {i.e.}, $\sigma_{E}(\vec{r}) = 0$. For the constraints, we choose an adiabatic wall. We consider two situations. In the first, the wall extends from left to right in a zigzag manner. Its position is given by $w_{1}(x) = h/2 - h \mathcal{H}(x - L/2)$ where $\mathcal{H}(x)$ is the Heaviside function. We refer to this constraint as \textit{vertical} (Figure \ref{fig:scheme-heat-flux}b). In the second, the wall is a straight line with a slope $k$,   {i.e.}, $w_{2}(x) = k(x - L/2)$. We refer to this constraint as \textit{linear} (Figure \ref{fig:scheme-heat-flux}c). Both constraints are fixed at $x=L/2$, so that the subsystems are symmetric in shape. Furthermore, they are chosen to ensure a non-zero heat flow in each subsystem. In other words, each subsystem is always in contact with both boundaries. 

For the unconstrained case, the temperature profile only depends on $x$. Solving Equation (\ref{T_r}) (which is now $\partial_{x}^{2}T(x)=0$) with boundary conditions $T(x=0) = T_{1}$ and $T(x=L) = T_{0}$, we have
\begin{equation}
T(x) = \dfrac{T_{0}-T_{1}}{L} x + T_{1}. 
\end{equation}
The energy density is then 
\begin{equation}
\epsilon = \epsilon_{0} \ddfrac{\til{T_{1}} - 1}{\ln \til{T_{1}}}. 
\end{equation}

Since we choose $T_{1} \geq T_{0}$, the heat flow passes through the system from left to right.~The~unit normal vector of the left (right) boundary is $\hat{n} = (-1,0)$ ($\hat{n} = (1,0)$). The local heat flux is $\vec{J}(\vec{r}) = -k(\partial_{x} T(\vec{r}), \partial_{y} T(\vec{r}))$. Hence, the heat flow going through the system can be calculated using either of the following expressions, 
\begin{equation}
\begin{aligned}
&\Phi_{in} = k\int_{Z} dz\int_{Y} \partial_{x}T(x,y)\mid_{x=0} dy, \\
&\Phi_{out} = k\int_{Z} dz\int_{Y} \partial_{x}T(x,y)\mid_{x=L} dy, 
\label{phi_heat}
\end{aligned}
\end{equation} 
which gives,
\begin{equation}
J_{U} = \dfrac{kAT_{0}}{L}(\til{T_{1}}-1). 
\label{eq:flux_HF}
\end{equation}

For the constrained system, the temperature profile depends also on the $y$-coordinate. It satisfies the equation $\nabla^{2}T(x,y) = 0$ with the following boundary conditions, 
\vspace{12pt}
\begin{equation}
\begin{cases} 
 T_{1}(0,y) = T_{1}, & T_{2}(0,y) = T_{1}, \\
 T_{1}(L,y) = T_{0} , & T_{2}(L,y) = T_{0}, \\
 \partial_{\hat{\mathbf{n}}}T_{1}(x,y)\mid_{y = w(x)} = 0, & \partial_{\hat{\mathbf{n}}}T_{2}(x,y)\mid_{y = w(x)} = 0,\\
 \partial_{y}T_{1}(x,y)\mid_{y = -H/2} = 0, & \partial_{y}T_{2}(x,y)\mid_{y = H/2}= 0.
 \end{cases}
 \label{bc}
\end{equation}
At $x= 0$ and $L$, the system is in contact with the plates. This is represented by the Dirichlet boundary conditions. In addition, since the constraint is an adiabatic wall, we have Neumann boundary conditions at $w_i(x), i=1,2$. Finally, at the boundaries far away from the constraint, we expect the effect of the constraint to diminish. In other words, at $y=\pm H/2$, we assume that the heat fluxes are parallel to the $x$-axis. To ensure this, we need to set $H/2 \gg w_i(L)$ and $H/2 \gg w_i(0)$ for $i=1, 2$. 

\begin{figure}[htb!]
\centering
\includegraphics[scale=0.5]{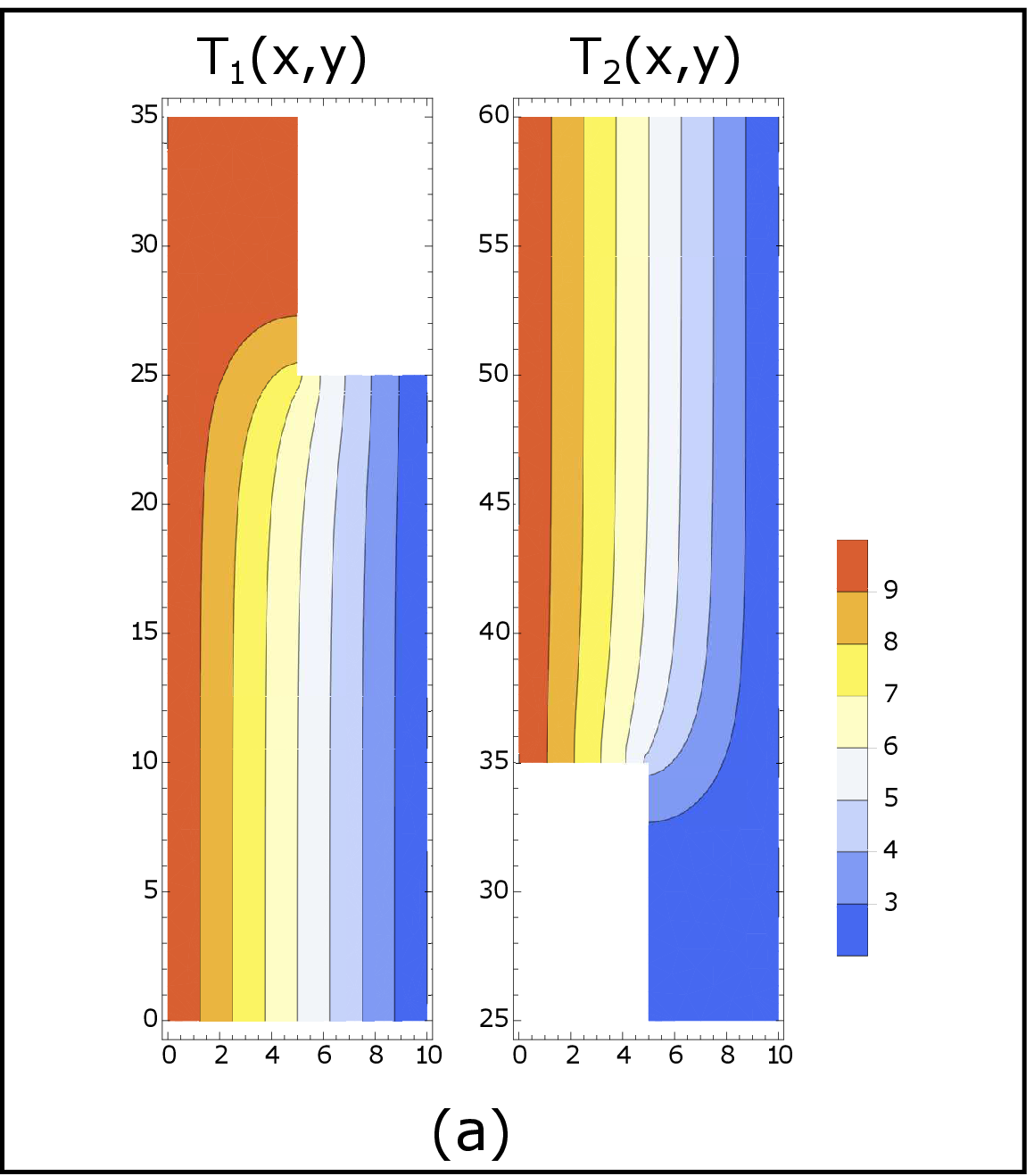} \hspace{12pt}
\includegraphics[scale=0.5]{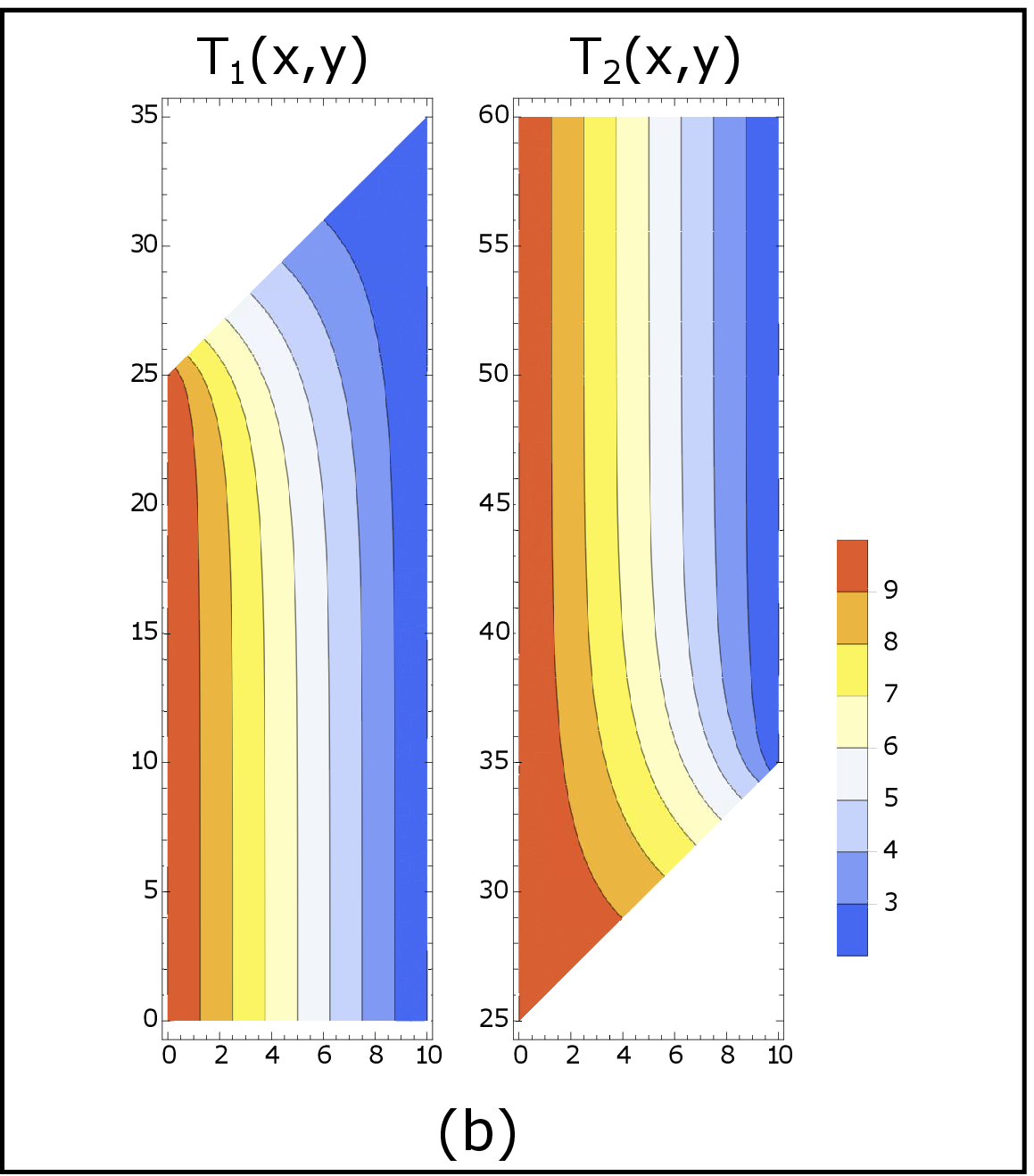}
\caption{Contour plots of temperature profiles: (\textbf{a}) results of a vertical constraint; and  (\textbf{b}) results of a linear constraint. In both figures, the temperatures at the boundaries are $T_{1}=10, T_{0}=2$. The size of the system is $L = 10$ and $H=60$. For the vertical constraint, the height of the wall is $h=10$. For the linear constraint, the slope of the wall is $k=1$.}
\label{Txy}
\end{figure}

The temperature profiles are obtained numerically using the finite element method. In this method, the system is separated into small domains called \textit{mesh} and the function is approximated using polynomials \cite{reddy2005introduction}. Examples of the contour plot of temperature profiles are shown in Figure \ref{Txy}. 

After obtaining temperature profiles, the stored energy density is calculated using Equation (\ref{e_intro}). The total heat flow is obtained using either the left or right boundary according to Equation (\ref{phi_heat}), 
\begin{equation}
J_{tot}=J_{U_1}+J_{U_2} = kZ\int_{-H/2}^{y^*} \partial_{x}T_{1}(x,y)\mid_{x=0} dy + kZ\int_{y^*}^{H/2} \partial_{x}T_{2}(x,y)\mid_{x=0} dy, 
\end{equation}
where $y^* = w_{1}(x=0)$ ($w_{2}(x=0)$) for the vertical (linear) constraint.

\begin{figure}[htb!]
\centering
\includegraphics[scale=0.25]{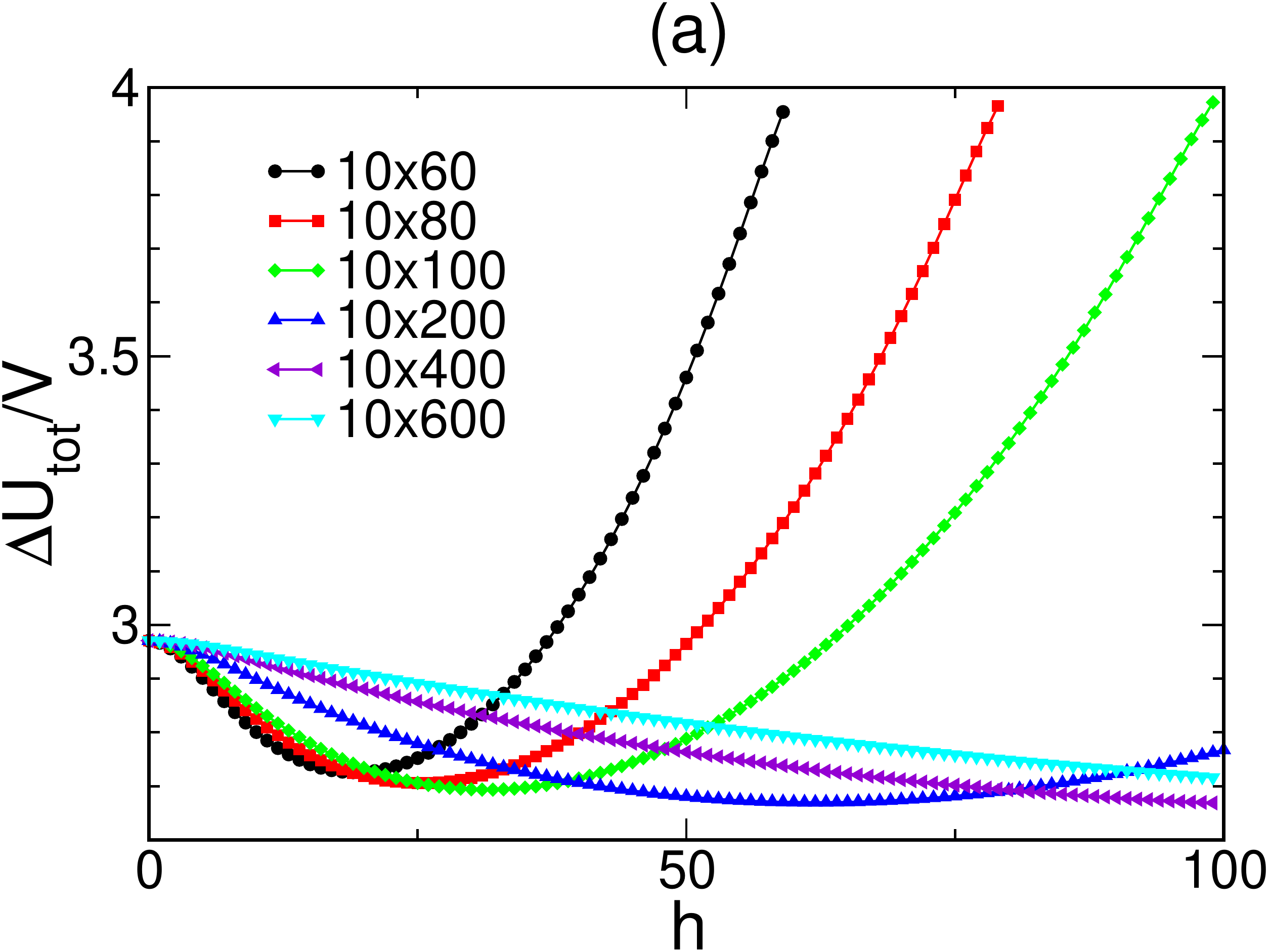}
\includegraphics[scale=0.25]{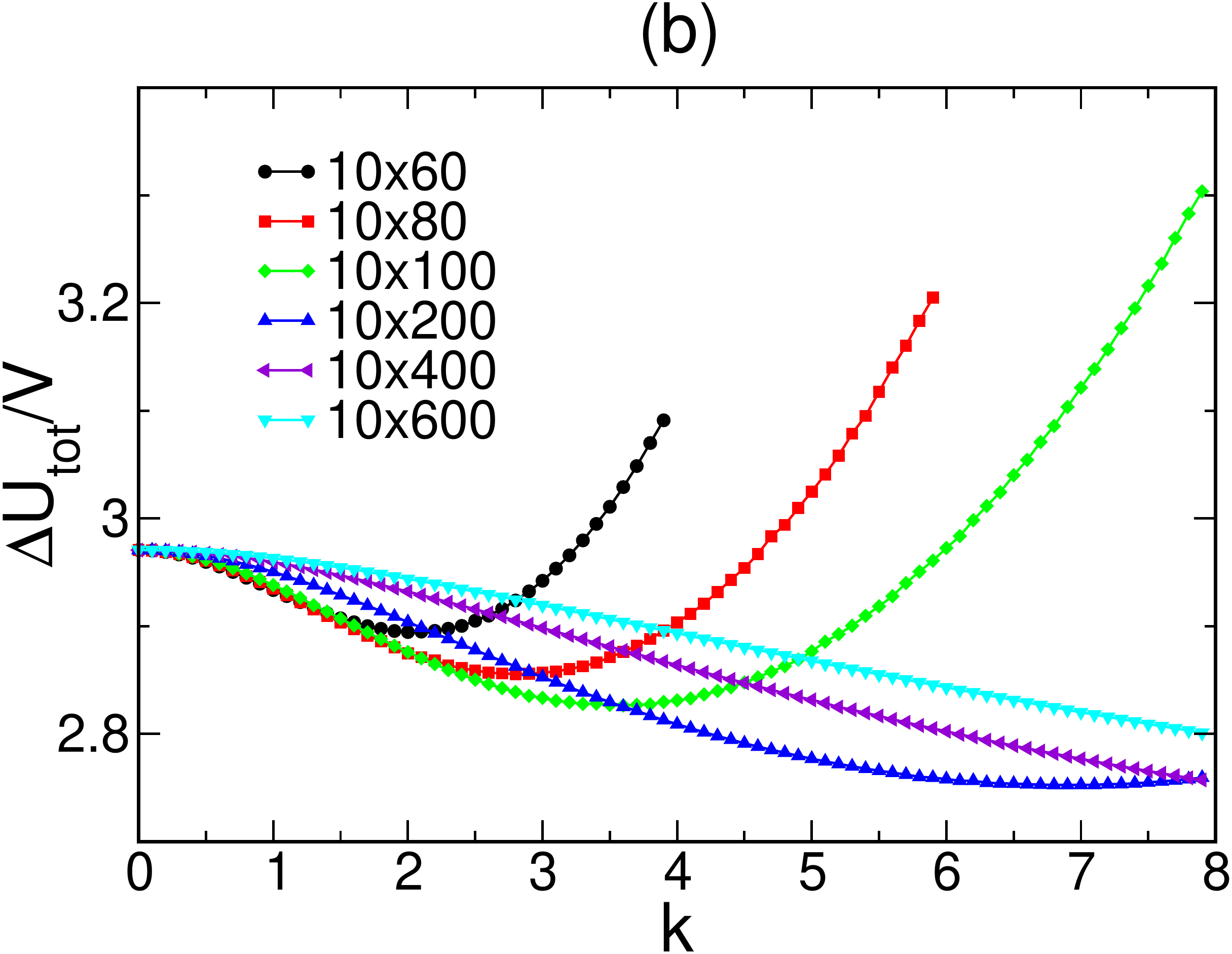}\\
\vspace{0.5cm}
\includegraphics[scale=0.25]{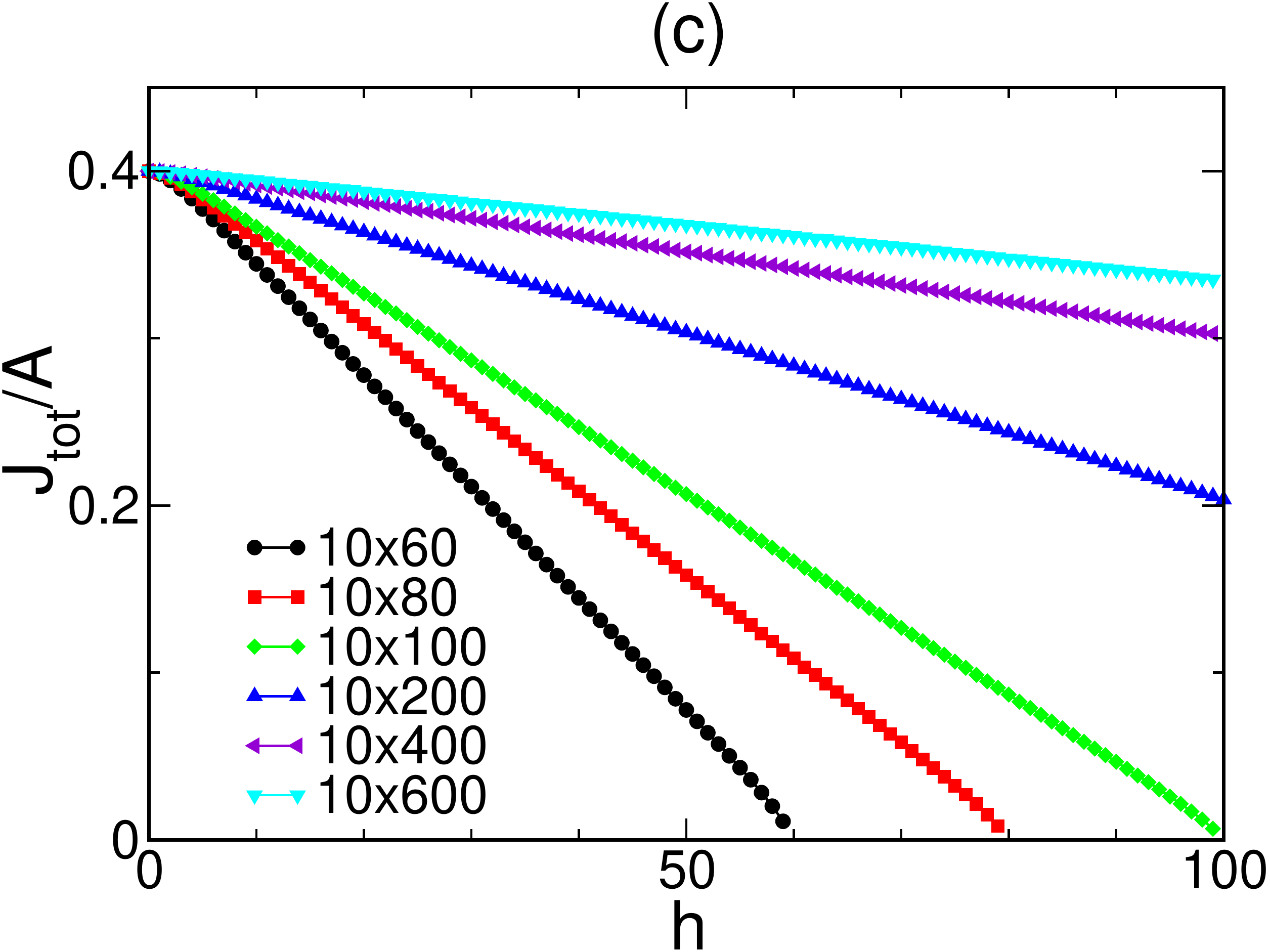}
\includegraphics[scale=0.25]{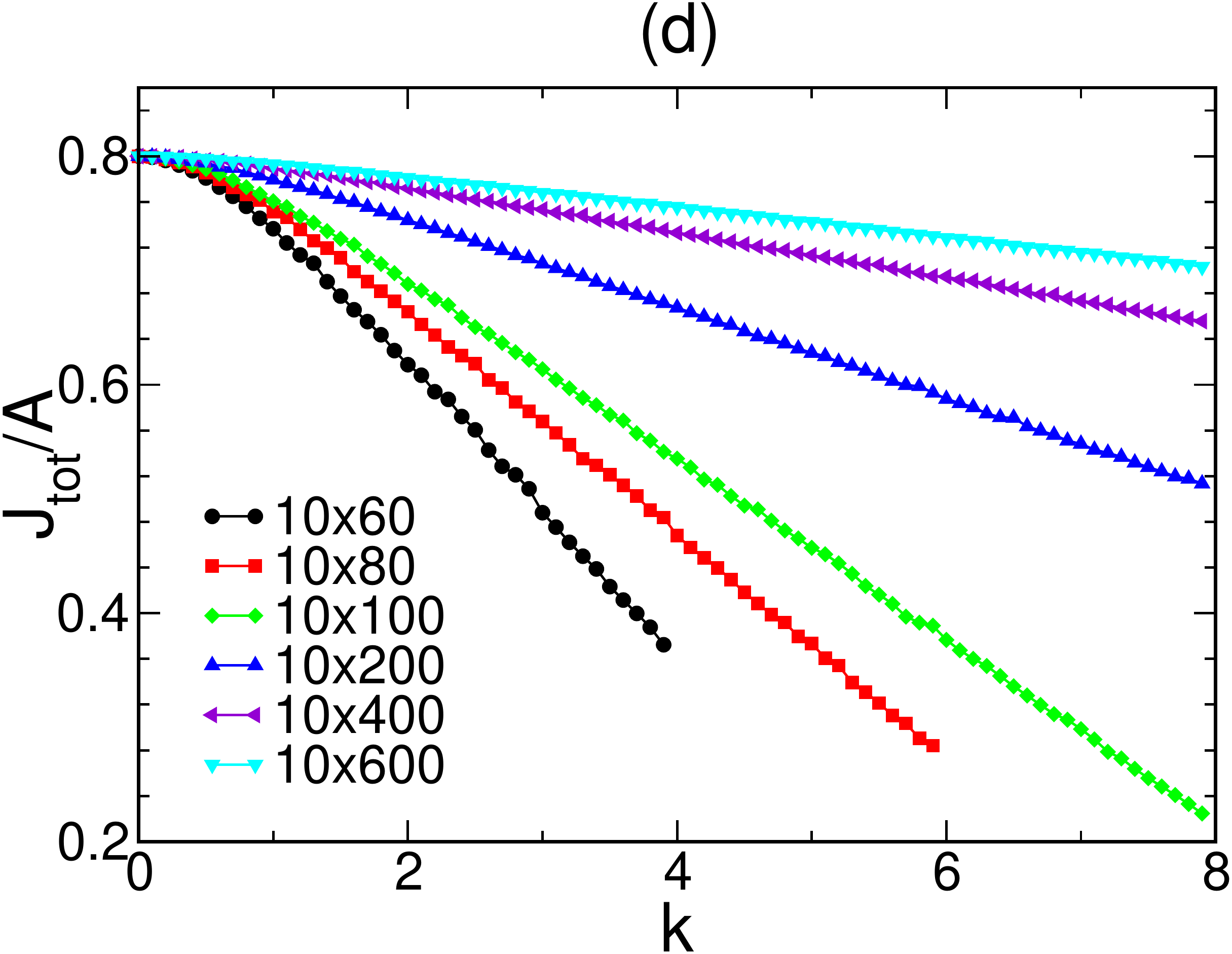}\\
\vspace{0.5cm}
\includegraphics[scale=0.25]{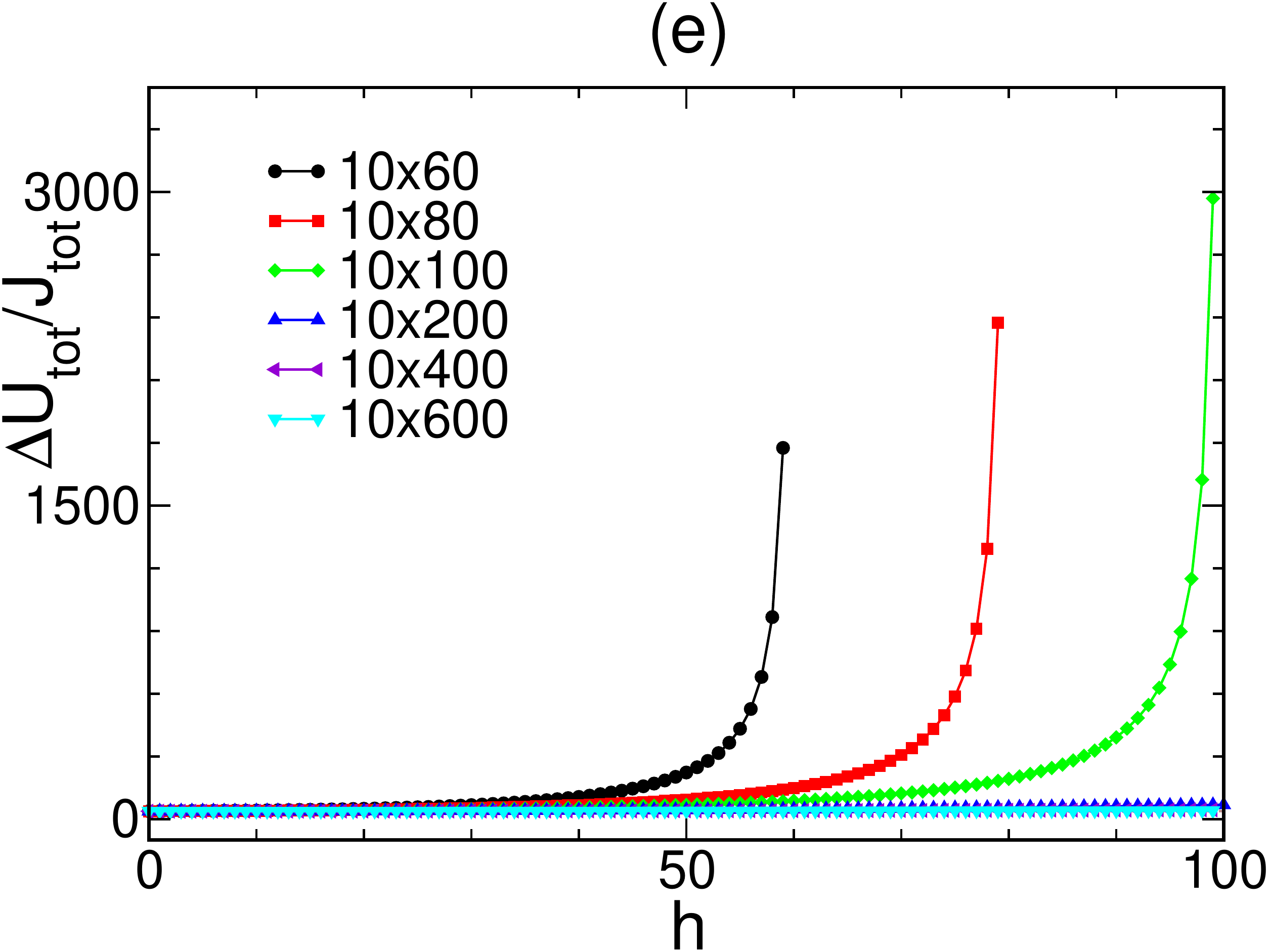}
\includegraphics[scale=0.25]{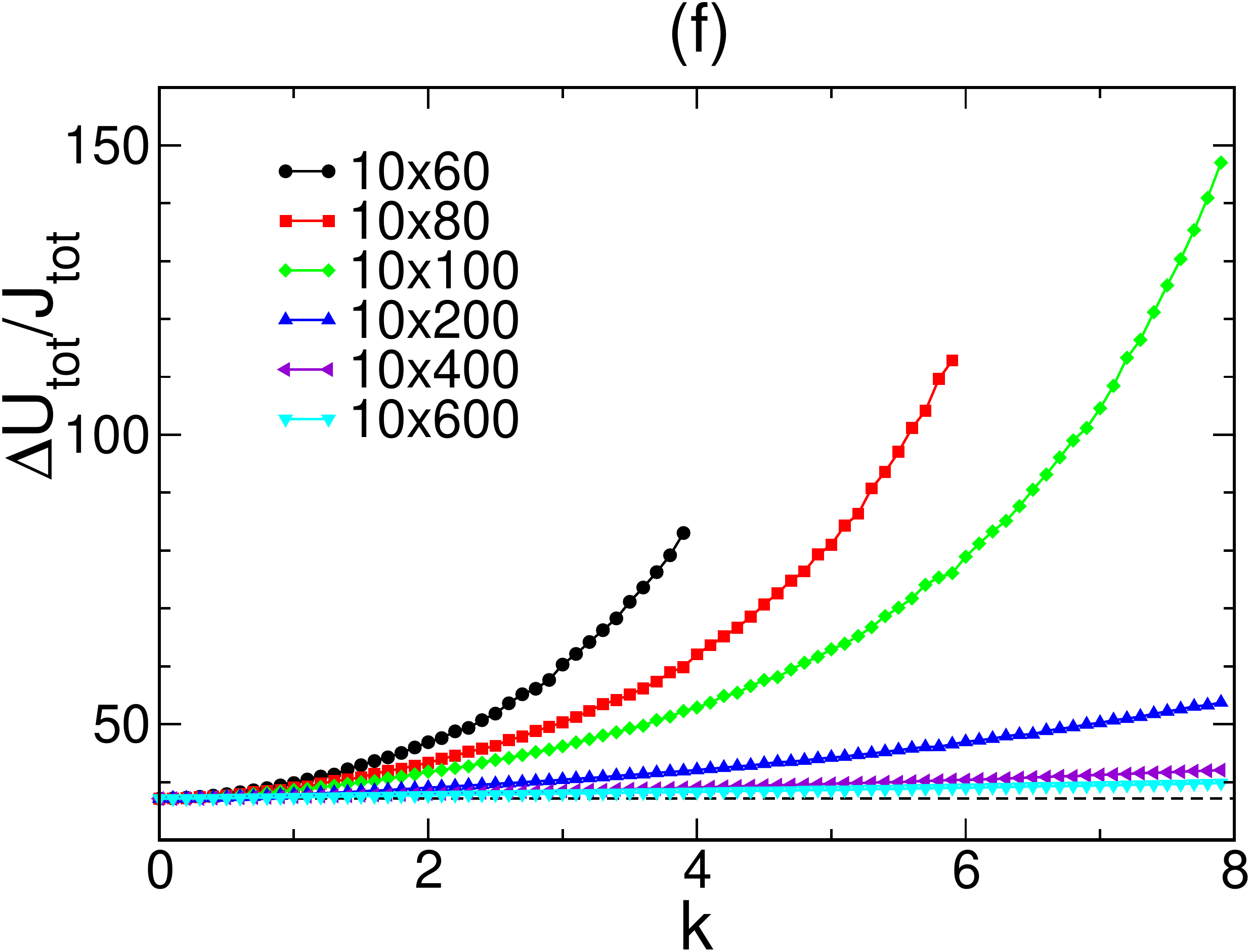}
\caption{Plots of total energy storage per volume $\Delta U_{tot}/V = \Delta (U_{1} + U_{2})/V$, total out-going heat flow per area $J_{tot}/A= (J_{U_1} + J_{U_2})/A$ and their ratio $\mathcal{T}_{1|2} = \Delta U_{tot}/J_{tot} = \Delta (U_{1} + U_{2})/(J_{U_1} + J_{U_2})$: results for vertical constraints   (\textbf{a},\textbf{c},\textbf{e}); and results for linear constraints   (\textbf{b},\textbf{d},\textbf{f}). Each panel is evaluated for six  different system sizes of a fixed $L=10$ and $H=60,80,100,200,400$ and $600$.}
\label{F}
\end{figure}

For both constraints, we study  $\Delta U_{tot}/V$, $J_{tot}/A$ and $\mathcal{T}_{1|2}$ at different parameters $h$ and $k$ with different system sizes (see Figure \ref{F}). 
In all cases, we find $\mathcal{T}_{1|2}(h) \geq \mathcal{T}_{1|2}(0)$ and $\mathcal{T}_{1|2}(k) \geq \mathcal{T}_{1|2}(0)$. If $h=0$ and $k=0$, the system is separated into identical subsystems and $\mathcal{T}_{1|2}(0) = \mathcal{T}$. Hence, $\mathcal{T} \leq \mathcal{T}_{1|2}$ for all these systems. 

\subsection{Matter Flow}

\begin{figure}
\centering
\includegraphics[scale=0.6]{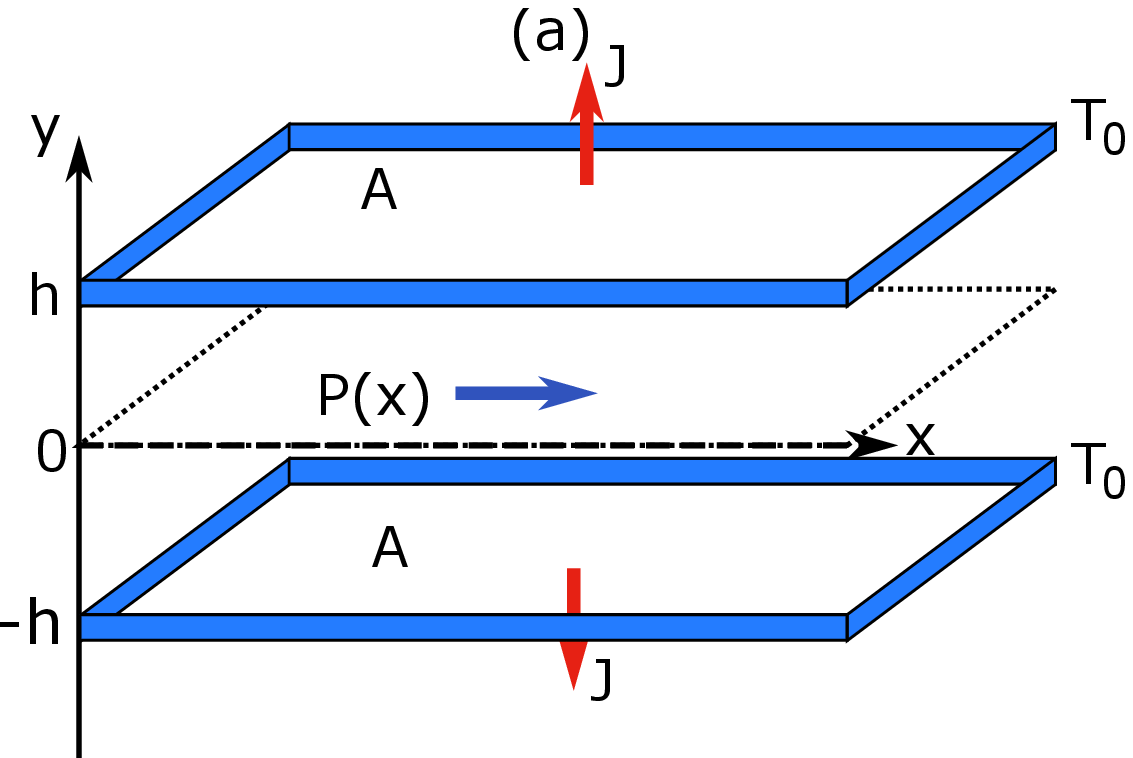}
\includegraphics[scale=0.6]{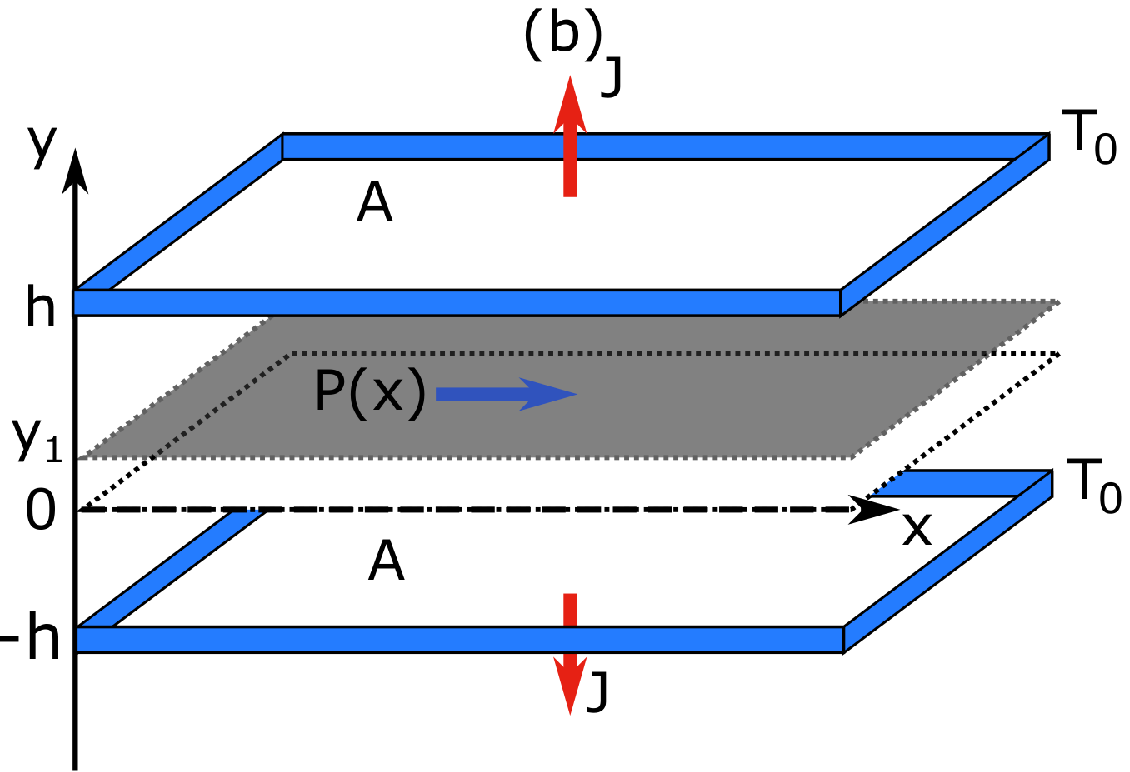}
\caption{Schemes of \textbf{(a)} unconstrained and \textbf{(b)} constrained Poiseuille flow. The system is bounded by two plates with a fixed temperature $T_{0}$ and area A that are placed at $y=\pm h$. A constant pressure gradient is applied across the system. In (\textbf{b}), the system is divided by an adiabatic slip wall placed at $y=y_1$.}
\label{scheme-pois}
\end{figure}

In   Case (iii), the ideal gas is flowing between two parallel walls located at $y = \pm h$ (see Figure \ref{scheme-pois}). The flow is assumed to be laminar and the fluid incompressible. It is driven by a constant pressure gradient along the $x$-axis, $\partial_{x}P(x) = -\mathbb{P}$. Such a flow is known as the Poiseuille flow \cite{graebel2007advanced}. Both walls are kept at temperature $T_{0}$. An adiabatic slip wall is introduced as the constraint into the system. It is placed at $y=y_{1}$ with $0\le y_1 \le 1$.

In the steady state, the velocity profile and the temperature profile can be obtained from the Navier--Stokes equation. We note that, due to the presence of the external pressure gradient and since the mass density of the incompressible fluid is homogeneous $\rho = \rho_{0}$, the energy density is not constant throughout the system. We first obtain the velocity profile $\vec{v} = (v(y),0)$ from 
\begin{equation}
\dfrac{\partial^{2}v(y)}{\partial y^{2}} = -\dfrac{\mathbb{P}}{\mu}, 
\end{equation}
where $\mu$ is the viscosity. Given the non-slip conditions at the boundaries $v(\pm h) = 0$, we find
\begin{equation}
v(y)=\dfrac{\mathbb{P}}{2\mu}(h^{2}-y^{2}). 
\end{equation}

Secondly, from the momentum equation, we obtain the dissipation density function $\phi = \mu(\partial_{y}v)^{2} = \mathbb{P}^{2}y^{2}/\mu$. The dissipation density function governs the rate at which the mechanical energy of the flow is converted to heat. The out-going heat flow $J_{U}$ is given by, 
\begin{equation}
J_{U} = A\int_{-h}^{h}\phi dy = V\dfrac{\mathbb{P}^{2}h^{2}}{3\mu}, 
\label{eq:flux_matt}
\end{equation}
where $A$ is the area of the plates and $V = A \times 2h$ is the volume of the system. 
Moreover, we assume that the heat transfer obeys the Fourier's law. From (Equation (\ref{T_r})), 
\begin{equation}
-k\dfrac{\partial^{2}T}{\partial y^{2}} = \phi, 
\end{equation}
together with boundary conditions $T_{\pm h} = T_{0}$, we obtain the temperature profile, 

\begin{equation}
T(y) = \dfrac{\mathbb{P}^{2}}{12\mu k}(h^{4}-y^{4}) + T_{0}. 
\end{equation}

Finally, we assume that the internal energy locally obeys the ideal gas law given by Equation (\ref{app:eps}). The total energy of the system consists of the kinetic energy and the internal energy, 
\begin{align}
E_{k} &= \dfrac{\rho_{0} A}{2} \int_{-h}^{h} v^{2}(y)dy 
= V\dfrac{\rho_0 \mathbb{P}^{2} h^{4}}{15\mu^{2}}, \label{pois_epsilon1} \\
E_{u} &= \dfrac{3}{2} A n_0k_B\int_{-h}^{h} T(y) dy 
= V \dfrac{n_{0} k_B \mathbb{P}^{2} h^{4}}{10\mu k} + \dfrac{3}{2} V n_{0} k_BT_0, 
\label{pois_epsilon2}
\end{align}
where the number density $n_0 = \rho_0/m$, with $m$   the mass of a single atom or molecule. Combining Equations (\ref{pois_epsilon1}), (\ref{pois_epsilon2}), and (\ref{eq:flux_matt}), we obtain
\begin{equation}
\mathcal{T}= \dfrac{\Delta U}{J_{U}} = \dfrac{E_{k} + E_{u} - E_{u_{0}}}{J_{U}} = n_{0} h^{2}\left(\dfrac{m}{5\mu} + \dfrac{3k_B}{10k}\right).
\end{equation}

For the constrained system, additional boundary conditions are $d v_{1}/dy\mid_{y=y_{1}} = 0$, $d v_{2}/dy\mid_{y=y_{1}} = 0$, $d T_{1}/dy\mid_{y=y_{1}} = 0$ and $d T_{2}/dy\mid_{y=y_{1}} = 0$.
Following the same method, we find the velocity profiles as 
\begin{align}
v_{1} &= \dfrac{\mathbb{P}}{2\mu} \big( (h-y_{1})^{2} - (y-y_{1})^{2} \big), \\
v_{2} &= \dfrac{\mathbb{P}}{2\mu} \big( (h+y_{1})^{2} - (y-y_{1})^{2} \big), 
\end{align}
and the temperature profiles as 
\begin{align}
T_{1}(y) &= \dfrac{\mathbb{P}^{2}}{12\mu k}\big( (h-y_{1})^{4} - (y-y_{1})^{4} \big) + T_{0}, \\
T_{2}(y) &= \dfrac{\mathbb{P}^{2}}{12\mu k}\big( (h+y_{1})^{4} - (y-y_{1})^{4} \big) + T_{0}. 
\end{align}
From these equations, we obtain
\begin{equation}
\mathcal{T}_{1|2} \equiv \dfrac{\Delta (U_1+U_2)}{(J_{U_1}+J_{U_2})} = n_0\dfrac{\big( (h-y_{1})^{5} + (h+y_{1})^{5} \big)}{\big( (h-y_{1})^{3} + (h+y_{1})^{3} \big)} \left(\dfrac{m}{5\mu} + \dfrac{3k_B}{10k}\right). 
\end{equation}

Comparing $\mathcal{T}$ and $\mathcal{T}_{1|2}$, the relation reduces to 
\begin{equation}
h^{2} \sim \dfrac{\big( (h-y_{1})^{5} + (h+y_{1})^{5} \big)}{\big( (h-y_{1})^{3} + (h+y_{1})^{3} \big)}. 
\end{equation}
Analysis shows that $\mathcal{T} \leq \mathcal{T}_{1|2}$. 

\subsection{Energy Density as Function of Heat Flow}
It is interesting to note that, for all the above studied models, the steady state energy density $\epsilon$ is a product of the equilibrium energy density $\epsilon_0$ and a dimensionless function of the heat flow $J_U$. For the ideal gas system with a homogeneous energy supply, where $J_U = 2LA\lambda$, $\epsilon$ can be written as (compare Equation (\ref{dia_epsilon})) 
\begin{equation}
\epsilon = \ddfrac{\epsilon_{0}\sqrt{\frac{L}{AkT_0}J_U\left(\frac{L}{AkT_0}J_U+4\right)}}{4\arctanh(\sqrt{\frac{L}{AkT_0}J_U/\left(\frac{L}{AkT_0}J_U+4\right)}}. 
\label{dia_epsilon_flux}
\end{equation}
Next, for the heat flow model, with $J_{U} = (T_{1}-T_{0})Ak/L$ (compare Equation (\ref{eq:flux_HF})), the steady state energy density can be expressed as 
\begin{equation}
\epsilon = \ddfrac{\epsilon_{0}\dfrac{L}{AkT_{0}}J_{U}}{\ln \big( \ddfrac{L}{AkT_{0}}J_{U} + 1 \big)}. 
\end{equation}
Lastly, for the matter flow model, with $J_{U} = 2A\mathbb{P}^{2}h^{3}/3\mu$ (compare Equation (\ref{eq:flux_matt})), the steady state internal energy density can be expressed as (compare Equation (\ref{pois_epsilon2})), 
\begin{equation}
\epsilon = \epsilon_{0} + \dfrac{1}{10} \dfrac{J_{U}h}{Ak T_{0}} \times \dfrac{3}{2}n_{0}k_BT_{0} 
 = \epsilon_{0} (1 + \dfrac{1}{10} \cdot \dfrac{h}{AkT_{0}}J_{U}). 
\end{equation} 
Thus, in all studied steady states, we find $U=U_0*f(J_UL/(AkT_0))$. 

\section{Conclusion}
We use the ideal gas model with  three different energy delivery methods to test the hypothesis that $\Delta U/J_{U}$ is minimized in steady states.  The results in all models confirm that $\Delta U/J_{U} \leq \Delta (U_{1} + U_{2})/(J_{U_1} +J_{U_2})$. 

Further, in all studied steady states, we find $U=U_0*f(J_UL/(AkT_0))$ and therefore $J_U$ is a parameter of NESS. By making a Legendre transform of $U$ with respect to $J_U$, we get an analog of the Helmholtz free energy for NESS, especially since   $J_U/T_0$ is the entropy flow leaving the system through the wall at temperature $T_0$. 
We introduce a quantity $U-(dU/dJ_U)J_U=U^*$, which we call the \textit{embedded energy}, since it is the stored energy minus the outflow of energy in the characteristic time $\tau=dU/dJ_U$. Thus, $U^*$ represents the part of the energy that must stay in the system for all times to keep the outflow of energy, while $\tau J_U$ is the energy that constantly flows through the system in time $\tau$. 
\acknowledgements
The work of Y.Z. was partially supported by the Polish National Science Centre (Harmonia Grant No. 2015/18/M/ST3/00403). The work of K.G. was supported by the Polish National Science Centre (Sonata Bis Grant No. SONATA BIS 2017/26/E/ST4/00041).
The work of R.H. was supported by the Polish National Science Centre (Maestro UMO-2016/22/A/ST4/00017). 
\appendix

\section{Appendix}
\label{appendix:A}
Here, we provide the calculation of the energy density $\epsilon$ from the temperature profile $T(\vec{r})$ with $\vec{r} = (x,y,z)$. 

The system has a fixed number of particles $N$ and a fixed volume $V$. It obeys the ideal gas law, 
\begin{align}
P &= n k_{B}T, \\
\epsilon &= \dfrac{3}{2}n k_{B}T, 
\end{align}
where $P$ is the pressure, $n = N/V$ is the particle number density, $k_{B}$ is the Boltzmann constant, and $\epsilon$ is the (internal) energy density. For steady states in   Cases (i) and (ii), we assume that the pressure (and hence the energy density) is homogeneous across the system, that is, 
\begin{equation}
P = n(\vec{r}) k_{B} T(\vec{r}), 
\end{equation}
\begin{equation} \label{app:eps}
\epsilon = \dfrac{3}{2} n(\vec{r}) k_{B} T(\vec{r}). 
\end{equation}

The energy density can be obtained by observing that
\begin{equation}
\epsilon \int_{V} \dfrac{d^3r}{T(\vec{r})} = \dfrac{3}{2}k_{B} \int_{V} d^3r n(\vec{r}) = \dfrac{3}{2}k_{B}N = \dfrac{\epsilon_{0}}{T_{0}}V, 
\end{equation}
where $n_{0}$ is the number density at equilibrium and we used $n_{0} V = \int_{V}d^3rn(\vec{r}) = N$. We denote the equilibrium variables with subscript $0$. The energy density is thus 
\begin{equation}
\epsilon = \dfrac{\epsilon_{0}}{T_{0}}\ddfrac{V}{\int_{V}\dfrac{d^3r}{T(\vec{r})}}. 
\label{e}
\end{equation}

When introducing the constraints, the subsystems are separated in such a way that $N_{i}/V_{i} = n_{0}$. For each subsystem, we have $\int_{V_{i}} d^3rn_{i}(\vec{r}) = N_{i}$ and, thus, 
\begin{equation}
\epsilon_{i} \int_{V_{i}} \dfrac{d^3r}{T_{i}(\vec{r})} = \dfrac{3}{2}k_{B} \int_{V_{i}} d^3r n(\vec{r}) = \dfrac{3}{2}k_{B}N_{i} = \dfrac{\epsilon_{0}}{T_{0}}V_{i}. 
\end{equation}
Therefore, the expression for $\epsilon_{i}$ is of the same form as $\epsilon$, 
\begin{equation}
\epsilon_{i} = \dfrac{\epsilon_{0}}{T_{0}}\ddfrac{V_{i}}{\int_{V_{i}}\dfrac{d^3r}{T_{i}(\vec{r})}}. 
\label{e_sub}
\end{equation}


\begin{thebibliography}{-------}
\providecommand{\natexlab}[1]{#1}

\bibitem[Ho{\l}yst and Poniewierski(2012)]{holyst2012thermodynamics}
Ho{\l}yst, R.; Poniewierski, A.
\newblock {\em Thermodynamics for Chemists, Physicists and Engineers};
{Springer:  Dordrecht,  The~Netherlands},  2012.

\bibitem[Martyushev(2013)]{martyushev2013entropy}
Martyushev, L.M.
\newblock Entropy and entropy production: Old misconceptions and new
  breakthroughs.
\newblock {\em Entropy} {\bf 2013}, {\em 15},~1152--1170.

\bibitem[Oono and Paniconi(1998)]{oono1998steady}
Oono, Y.; Paniconi, M.
\newblock Steady state thermodynamics.
\newblock {\em {Prog. Theor. Phys. Supp.}} {\bf 1998}, {\em
  130},~29--44.

\bibitem[Dickman and Zia(2018)]{dickman2018driven}
Dickman, R.; Zia, {R.K.P}.
\newblock Driven Widom-Rowlinson lattice gas.
\newblock {\em {Phys. Rev. E}} {\bf 2018}, {\em 97},~062126.

\bibitem[Bartlett and Virgo(2016)]{bartlett2016maximum}
Bartlett, S.; Virgo, N.
\newblock {Maximum entropy production is not a steady state
  attractor for 2D fluid convection}.
\newblock {\em Entropy} {\bf 2016}, {\em 18},~431.

\bibitem[Morriss and Truant(2012)]{morriss2012deterministic}
Morriss, G.P.; Truant, D.
\newblock Deterministic thermal reservoirs.
\newblock {\em Entropy} {\bf 2012}, {\em 14},~1011--1027.

\bibitem[Holubec \em{et~al.}(2017)Holubec, Ryabov, Yaghoubi, Varga, Khodaee,
  Foulaadvand, and Chvosta]{holubec2017thermal}
Holubec, V.; Ryabov, A.; Yaghoubi, M.H.; Varga, M.; Khodaee, A.; Foulaadvand,
  M.E.; Chvosta, P.
\newblock Thermal ratchet effect in confining geometries.
\newblock {\em Entropy} {\bf 2017}, {\em 19},~119.

\bibitem[Jou \em{et~al.}(2010)Jou, Lebon, and
  Casas-V{\'a}zquez]{jou2010extended}
Jou, D.; Lebon, G.; Casas-V{\'a}zquez, J.
\newblock {\em {Extended Irreversible Thermodynamics}}, 4th ed.;
{Springer:  Dordrecht,  The~Netherlands},  2010.

\bibitem[Lebon \em{et~al.}(2008)Lebon, Jou, and
  Casas-V{\'a}zquez]{lebon2008understanding}
Lebon, G.; Jou, D.; Casas-V{\'a}zquez, J.
\newblock {\em {Understanding Non-Equilibrium Thermodynamics:
  Foundations, Applications, Frontiers}}; {Springer:  Dordrecht,  The~Netherlands},  2008.

\bibitem[Luzzi \em{et~al.}(2002)Luzzi, Vasconcellos, and
  Ramos]{luzzi2002predictive}
Luzzi, R.; Vasconcellos, {\'A}.R.; Ramos, J.G.
\newblock {\em {Predictive Statistical Mechanics: A
  Nonequilibrium Ensemble Formalism}}; {Springer:  Dordrecht,  The~Netherlands},  2002.

\bibitem[M{\"u}ller and Ruggeri(2013)]{muller2013rational}
M{\"u}ller, I.; Ruggeri, T.
\newblock {\em {Rational Extended Thermodynamics}}; Springer Science \& Business Media: Berlin/Heidelberg, Germany, 2013.

\bibitem[Eu(2013)]{eu2013nonequilibrium}
Eu, B.C.
\newblock {\em {Nonequilibrium Statistical Mechanics: Ensemble
  Method}}; Springer Science \& Business Media: Berlin/Heidelberg, Germany, 2013.

\bibitem[Ho{\l}yst \em{et~al.}(2019)Ho{\l}yst, Macio{\l}ek, Zhang, Litniewski,
  Knycha{\l}a, Kasprzak, and Banaszak]{Robert}
Ho{\l}yst, R.; Macio{\l}ek, A.; Zhang, Y.; Litniewski, M.; Knycha{\l}a, P.;
  Kasprzak, M.; Banaszak, M.
\newblock Flux and storage of energy in non-equilibrium, stationary states.
\newblock {\em Phys. Rev. E} {\bf 2019}, {\em 99},~042118.

\bibitem[Casas-V{\'a}zquez and Jou(2003)]{casas2003temperature}
Casas-V{\'a}zquez, J.; Jou, D.
\newblock {Temperature in non-equilibrium states: A review of
  open problems and current proposals}.
\newblock {\em Rep. Prog. Phys.} {\bf 2003}, {\em 66},~1937--2023.

\bibitem[Puglisi \em{et~al.}(2017)Puglisi, Sarracino, and
  Vulpiani]{puglisi2017temperature}
Puglisi, A.; Sarracino, A.; Vulpiani, A.
\newblock {Temperature in and out of equilibrium: A review of
  concepts, tools and attempts}.
\newblock {\em Phys. Rep.} {\bf 2017}, {\em 709},~1--60.

\bibitem[Reddy(2005)]{reddy2005introduction}
Reddy, J.
\newblock {\em An Introduction to the Finite Element Method}, 3rd ed.;
{McGraw-Hill Education}:  New York, NY, USA, 2005.

\bibitem[Graebel(2007)]{graebel2007advanced}
Graebel, W.
\newblock {\em Advanced Fluid {Mechanics}}; {Academic
  Press}: Cambridge, MA, USA,  2007.

\end{thebibliography}

\end{document}